\documentclass[aps,prl,twocolumn,superscriptaddress]{revtex4}

\usepackage{amsmath,amssymb,graphicx,xcolor,bm,soul}
\usepackage{comment}

\usepackage[british]{babel}
\usepackage{upgreek}
\usepackage{xfrac}
\usepackage{soul}
\usepackage{bm}

\definecolor{orange}{RGB}{235, 129, 0}
\usepackage[bookmarks=true,
            bookmarksnumbered=false,
            bookmarksopen=true,
            breaklinks=true,
            pdfborder={0 0 0},
            final=true,
            backref=none,
            colorlinks=true,
            linkcolor=blue,
            menucolor=blue,
            citecolor=blue,
            urlcolor=blue]{hyperref}
            
\begin{document}

\title{Nonlinear quantum optics with trion-polaritons in 2D monolayers: conventional and unconventional photon blockade}

\author{O. Kyriienko}
\affiliation{Department of Physics and Astronomy, University of Exeter, Stocker Road, Exeter EX4 4QL, UK}
\affiliation{Department of Nanophotonics and Metamaterials, ITMO University, St. Petersburg, 197101, Russia}

\author{D.~N. Krizhanovskii}
\affiliation{Department of Physics and Astronomy, The University of Sheffield, Sheffield, S3~7RH, United Kingdom}
\affiliation{Department of Nanophotonics and Metamaterials, ITMO University, St. Petersburg, 197101, Russia}

\author{I.~A. Shelykh}
\affiliation{Science Institute, University of Iceland, Dunhagi-3, IS-107 Reykjavik, Iceland}
\affiliation{Department of Nanophotonics and Metamaterials, ITMO University, St. Petersburg, 197101, Russia}

\begin{abstract}
We study a 2D system of trion-polaritons at the quantum level and demonstrate that for monolayer semiconductors they can exhibit a strongly nonlinear optical response. The effect is due to the composite nature of trion-based excitations resulting in their nontrivial quantum statistical properties, and enhanced phase space filling effects. We present the full quantum theory to describe the statistics of trion-polaritons, and demonstrate that the associated nonlinearity persists at the level of few quanta, where \emph{two qualitatively different regimes of photon antibunching} are present for \emph{weak} and \emph{strong} single photon-trion coupling. We find that single photon emission from trion-polaritons becomes experimentally feasible in state-of-the-art transition metal dichalcogenide (TMD) setups. This can foster the development of quantum polaritonics using 2D monolayers as a material platform.
\end{abstract}

\maketitle

\textit{Introduction.---}Exciton-polaritons are hybrid quasiparticles formed in optical microcavities in the regime of strong light-matter coupling. Their unique properties related to the composite nature lead to a dramatic enhancement of the nonlinear optical response and enable the observation of quantum collective phenomena at relatively high temperatures \cite{CarusottoCiutiRev,ByrnesRev}. Examples include the observation of polariton BEC and polariton lasing \cite{Kasprzak2006,Balili2007,Schneider2013}, topological defects such as solitons \cite{Amo2011,Amo2012,Hivet2012,Chana2015,Boulier2015,Gulevich-SciRep-2017} and quantized vortices \cite{Rubo2007,Lagoudakis2008,Tosi2012,Gao2018,Kwon2019}, and many others. Moreover, polariton systems can form a basis for creation of nanophotonic devices of the next generation, including optical logic gates and all-optical integrated circuits \cite{ShelykhReview,TaniaEO2013,Opala2019}.

For conventional GaAs \cite{Balili2007} and CdTe \cite{Kasprzak2006} systems polaritonic nonlinearities mainly stem from exciton-exciton scattering processes, governed by the Coulomb exchange between electrons and holes \cite{Ciuti1998,Tassone1999}. However, another important contribution comes from phase-space filling effects related to the composite nature of excitons \cite{CombescotReview,Shiau2019}, also known as saturation effects \cite{Tassone1999}. In GaAs these effects were shown to be negligible at moderate pump powers \cite{Tassone1999}, but can become significant at certain cases \cite{Brichkin2011}. In particular, they govern the transition from strong to the weak coupling regimes at large pump powers \cite{Butte2002,Bajoni2008} and give dominant impact to the nonlinear response for the systems with Frenkel excitons \cite{Daskalakis2014,Yagafarov2009,Betzold2009}.

One of the most promising platforms for polaritonics is represented by transition metal dichalcogenide (TMD) monolayers \cite{LiuMenon2015,Dufferwiel2017,Lundt2017,Schneider2018,Zhang2018}. A remarkable compatibility of these 2D materials with various semiconductor/dielectric platforms makes them promising for the development of various nanophotonic devices. TMD excitons have extremely large binding energies and oscillator strengths as compared to excitons in conventional semiconductors, thus dominating an optical response even at room temperature \cite{Wang2018}. Moreover, it is also important that optical spectra of TMD monolayers reveal very robust trion \cite{Courtade2017,Lundt2018} and biexciton \cite{Nagler2018} peaks, and the peaks connected to excited exciton states \cite{Chernikov2014,Zipfel2018,Han2018,Yong2019}. In the polariton regime, to date the exciton-based nonlinear energy shift \cite{Barachati2018,Shahnazaryan2017} together with dissipative nonlinearity coming from exciton-exciton annihilation \cite{Kumar2014} were reported, and enhancement of nonlinearity in the cooperative coupling regime \cite{Wild2018} was proposed. 

Recently, the quality improvement of optical microcavities has much prolonged the lifetime of exciton polaritons \cite{Nelsen2013}, thus allowing to observe the first signatures of entering the quantum regime \cite{Munoz-Matunano2019}. 
Current state-of-the-art is represented by weak antibunching of $0.95$ \cite{Munoz-Matunano2019,Delteil2019}. However, the possibility to obtain stronger antibunching is ultimately limited by insufficient value of the effective exciton-exciton interaction constant responsible for the Kerr-type nonlinearity \cite{Verger2006}. This hinders the associated development of quantum polaritonics, and qualitatively new ideas are needed in order to propose how one can increase dramatically the nonlinear response of the system on single quantum level. One potential solution is to use the effects of quantum interference in double pillar systems without increasing the interaction constant itself \cite{Liew2010,Bamba2011,Snijders2018,Flayac2017,Zhou2015,Shen2015}. Another way corresponds to enhancing the interaction in dipolariton systems \cite{Togan2018,Rosenberg2018,Kyriienko2014} or using Rydberg excitons \cite{Kazimierczuk2014}. In the present letter we consider an alternative way to reach the regime of the polariton blockade, which does not rely on structure modification and is well suited to the 2D material platform. 

The object under study is a trion-polariton system based on TMD flakes placed in a photonic microresonator. We show that quantum statistical properties of trions stemming from their composite structure crucially affect the mechanism of light-matter coupling, and result in highly nonlinear optical response. We build the full non-perturbative quantum theory of the phase-space filling effects, and consider both coherence properties and optical spectra. In particular, we find regions of parameters where unconventional \emph{saturation-based blockade} can be achieved, as well as describe conventional blockade attainable at large single photon-trion couplings. We show that in state-of-the-art TMD polariton structures the antibunching of $g^{(2)}(0)<0.1$ can be achieved. Plotting the optical spectrum of the system at increasing pump power, we also find the conditions for the transition between strong and weak light-matter coupling regimes. 
\begin{figure}
\includegraphics[width=1.\linewidth]{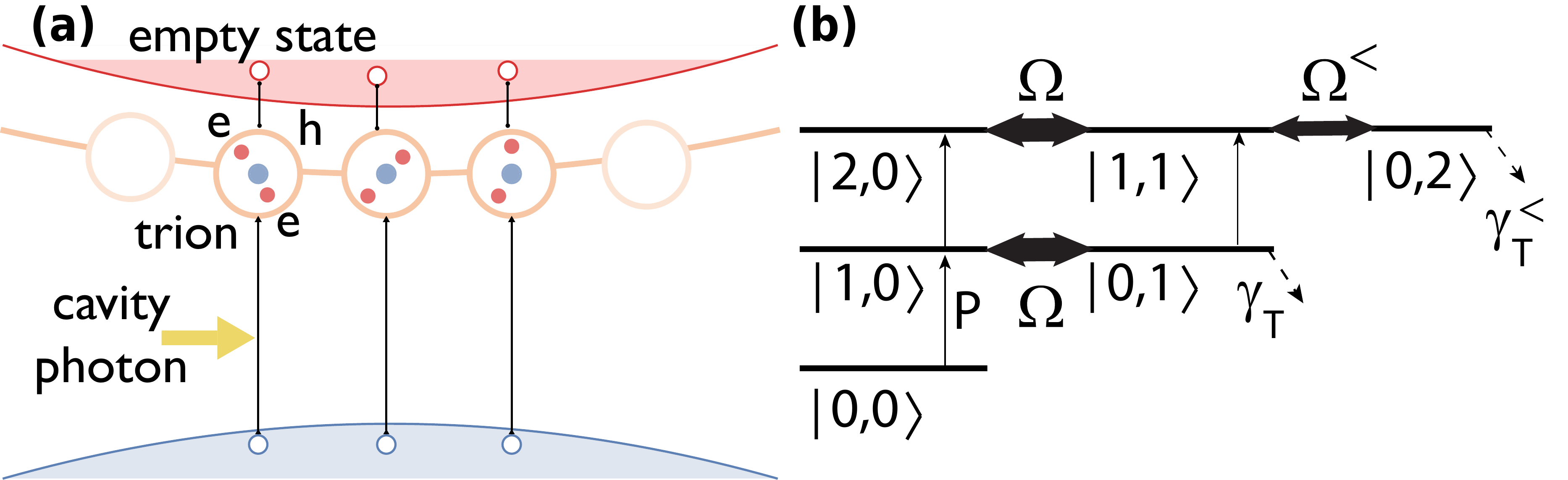}
\caption{\textbf{(a)} Schematic representation of trion-based excitations. A cavity photon creates an electron hole pair and captures an electron from the conduction band to form a trion complex. \textbf{(b)} Energy level scheme in photon-trion basis $|N_C, N_T\rangle$, where the direct excitation path $|0,0\rangle \rightarrow |1,0\rangle \rightarrow |2,0\rangle$ destructively interferes with trion-mediated path. The effective nonlinearity comes from the reduced coupling $\Omega^{<}$ in the presence of two trions.}
\label{fig:sketch}
\end{figure}

\textit{The model.---}We aim to build a quantum theory of trion-polaritons and study their optical properties. We start by considering a planar semiconductor which is initially electron doped. Optically excited electron-hole (e-h) pairs can interact with available free electrons  and form a bound trion state [Fig. \ref{fig:sketch}(a)]. The corresponding creation operator for this composite particle reads \cite{Ramon2003}
\begin{align}
\label{eq:T_singlet}
\hat{T}_{\mathbf{K},\mathbf{s}}^\dagger = \sum_{\mathbf{k}_1, \mathbf{k}_2, \mathbf{s}} \phi_{\mathbf{K}, \mathbf{k}_1, \mathbf{k}_2; \mathbf{s}} \hat{a}_{\mathbf{k}_1 ,s_1}^\dagger \hat{a}_{\mathbf{k}_2, s_2}^\dagger  \hat{b}_{\mathbf{K} -\mathbf{k}_1 - \mathbf{k}_2, s_3}^\dagger,
\end{align}
where the trion wavefunction $\phi_{\mathbf{K}, \mathbf{k}_1, \mathbf{k}_2; \mathbf{s}}$ is separated into a center-of-mass (CM) part with momentum $\mathbf{K}$, and the relative motion part described by the relative motion wavefunction $\phi_{\mathbf{k}_1,\mathbf{k}_2}$. Here, $\hat{a}_{\mathbf{k}, s_j}^\dagger$ and $\hat{b}_{\mathbf{k},s_j}$ are fermionic creation operators for electrons and holes, the indices $s_j$ correspond to the spins of individual fermions which define the spin configuration of a trion complex (denoted by the index $\mathbf{s}$), which can be a singlet or a triplet. In the following we consider only dominant trions in a singlet configuration, omitting spin indices for brevity. 

The full excitation process relies on taking an electron from the Fermi sea by photocreated e-h pair, where an empty electron state is left in the conduction band [Fig. \ref{fig:sketch}(a)]. This process can be conveniently described by the generation of the quasi-bosonic excitation in the system from the vacuum state $|\text{\O}\rangle$ corresponding to the Fermi sea, and we consider the low temperature case of a degenerate electron gas. The process is described by the excitation wavefunction \cite{Rapaport2000,Rapaport2001}
\begin{align}
\label{eq:B_K}
\hat{B}_{\mathbf{K}}^\dagger |\text{\O}\rangle = \frac{1}{\sqrt{N_s}} \sum_{\mathbf{k}} \hat{T}_{\mathbf{K} + \mathbf{k}}^\dagger \hat{a}_{\mathbf{k}} |\text{\O}\rangle ,
\end{align}
 where $N_s$ is a number of free electrons available for a trion creation. 
 As the excitation operator $\hat{B}_{\mathbf{K}}^\dagger$ contains four fermionic operators, it represents a composite boson with commutation relation $\Big[\hat{B}_{\mathbf{q}'}, \hat{B}_{\mathbf{q}}^\dagger \Big] = \delta_{\mathbf{q}',\mathbf{q}} - \hat{D}_{\mathbf{q}',\mathbf{q}}$, where operator $\hat{D}_{\mathbf{q}',\mathbf{q}}$ represents the deviation from bosonicity, and explicit form can be written straightforwardly using excitation operator in Eq.~\eqref{eq:B_K}. For tightly bound trions present in TMD monolayers the trion operators [Eq.~\eqref{eq:T_singlet}] can be treated as fermions, as their fine structure is only revealed at densities comparable to the inverse area of a trion $(\pi a_T^2)^{-1}$. In this case, the deviation operator corresponds to the fraction of trions formed out of Fermi sea, $\hat{D}_{\mathbf{q}',\mathbf{q}} = \sum_{\mathbf{k}} \hat{T}_{\mathbf{q}+\mathbf{k}}^\dagger \hat{T}_{\mathbf{q}'+\mathbf{k}}/N_s$. This makes trion-polariton excitations prone to the phase space filling effects and can lead to the saturation of light-matter coupling. At the same time, we remind that $\hat{B}_{\mathbf{K}}$ is not a bound state, and thus exhibits different statistics as compared to exciton-polaritons \cite{Brichkin2011,CombescotReview}, resembling more the case of intersubband polaritons \cite{DeLiberato2008,DeLiberato2012}.

The Hamiltonian of the system can be written as the sum of three terms, $\hat{\mathcal{H}} = \hat{\mathcal{H}}_0 + \hat{\mathcal{H}}_{\mathrm{coupl}} + \hat{\mathcal{H}}_{\mathrm{T-T}}$. Here, $\hat{\mathcal{H}}_0 $ describes non-interacting cavity photons, electrons, and trions. $\hat{\mathcal{H}}_{\mathrm{coupl}}$ describes the processes of light-matter coupling, and $\hat{\mathcal{H}}_{\mathrm{T-T}}$ describes Coulomb trion-trion scattering. In this paper we focus on the mechanism of nonlinearity stemming from the saturation effects related to Pauli exclusion principle, and neglect the latter term responsible for higher-order Coulomb effects.
The light-matter coupling Hamiltonian can be expressed in terms of the photonic operators $\hat{c}_{\mathbf{q}}$ and operators of quasibosonic excitations introduced in Eq.~\eqref{eq:B_K}, reading
\begin{align}
\label{eq:H_trion_int}
\hat{\mathcal{H}}_{\mathrm{coupl}} = \frac{\Omega}{2}\sum_{\mathbf{k}}\left( \hat{B}_{\mathbf{k}}^\dagger \hat{c}_{\mathbf{k}} +\hat{B}_{\mathbf{k}}\hat{c}_{\mathbf{k}}^\dagger\right),
\end{align}
where $\Omega = g_0\sqrt{N_s} \sum_{\mathbf{k}_1,\mathbf{k}_2} \phi_{\mathbf{k}_1,\mathbf{k}_2}$ is a Rabi energy corresponding to coupling between the collective trion mode and cavity photons (accounting for the trion localization), with $g_0$ being a valence-to-conduction band transition matrix element. 
\begin{figure}
\centering
\includegraphics[width=1.\linewidth]{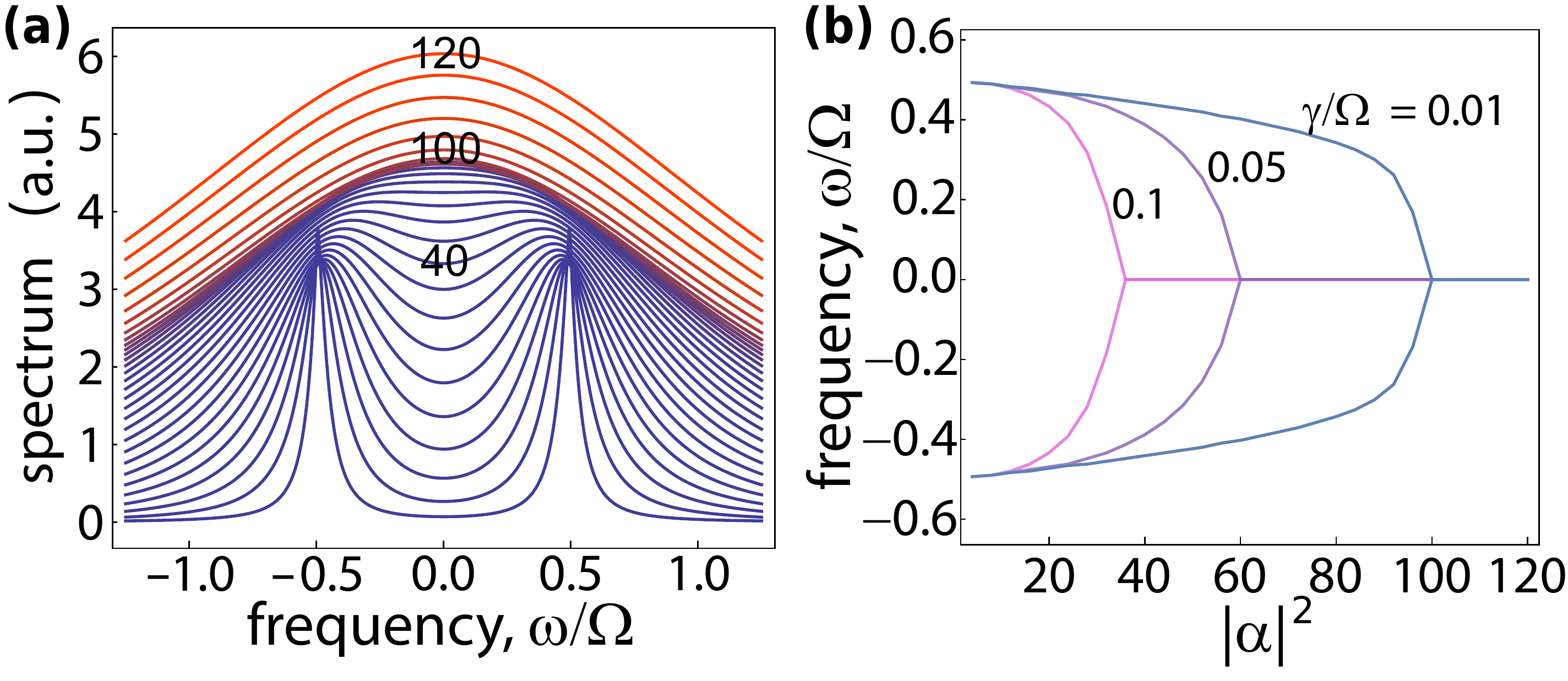}
\caption{\textbf{(a)} Absorption spectrum of a trion-polariton system for various photon numbers $N_C = |\alpha|^2$ being injected to the system by short coherent optical pulse. The parameters are fixed to $N_s = 100$ available electronic states, $\gamma_{c,T} = 0.05\Omega$, and we work at zero photon-trion detuning (values of $N_C = \{40, 100, 120\}$ are highlighted). \textbf{(b)} Spectral peak locations for lower and upper polariton are shown as a function of drive, and reveal the collapse of strong coupling.}
\label{fig:spectrum}
\end{figure}

The total number of excitations (photons plus trions) in the system is conserved. Thus, we can split the associated Hilbert-Fock space of the problem into separate manifolds corresponding to the different numbers of the excitations $N$, and then diagonalize each block separately. For this, let us define the matrix element $\mathcal{M}_{m',n'}^{m,n} := \langle m', n'| \hat{\mathcal{H}} | m, n \rangle$,  where $|N_C, N_T\rangle$ represents a state with $N_C$ photons (bosons) and $N_T$ quasi-bosonic trion excitations. The Hamiltonian describing $N = N_C + N_T -1$ particles is represented by the matrix
\begin{align}
\label{eq:H_N_matrix}
    \hat{H}_N = \begin{bmatrix} 
\mathcal{M}_{N_C, N_T -1}^{N_C, N_T -1} & \mathcal{M}_{N_C, N_T -1}^{N_C - 1, N_T} \\
\mathcal{M}_{N_C-1, N_T}^{N_C, N_T-1} & \mathcal{M}_{N_C-1, N_T}^{N_C-1, N_T} 
\end{bmatrix}.
\end{align}
The matrix elements entering the expression above can be calculated element-wise properly accounting for the composite nature of the particles. The diagonal elements read $\mathcal{M}_{N_C, N_T -1}^{N_C, N_T -1} = N_C \omega_{\mathrm{cav}} + \omega_{T} (N_T - 1)$,
%
%
where $\omega_{\mathrm{cav}}$ is an energy of a photonic cavity mode, and $\omega_{T}$ corresponds to the energy difference between electron and trion energies [SM].
\begin{figure}
\centering
\includegraphics[width=1.\linewidth]{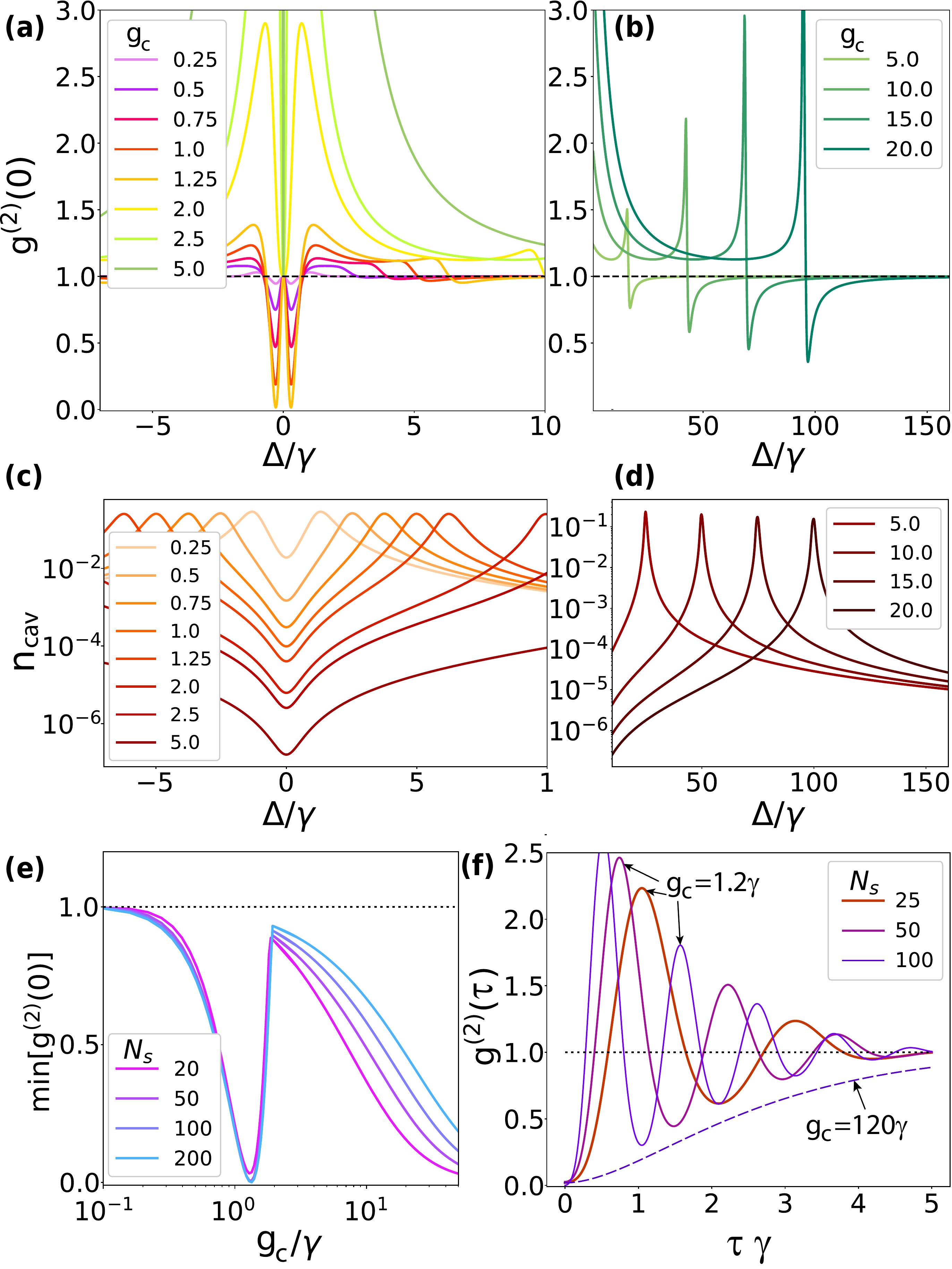}
\caption{Second order coherence at zero delay for trion-polariton system vs. different parameters. \textbf{(a, b)} $g^{(2)}(0)$ for cavity photons plotted as a function of pump detuning $\Delta = \omega_{\mathrm{cav}} - \omega_p$ for different values of single trion-photon coupling $g_c=\Omega/\sqrt{N_s}$. In \textbf{(a)} the unconventional blockade window at $\Delta \approx 0$ and small $g_c \approx \gamma$ is shown, and \textbf{(b)} shows the conventional regime at $\Delta \approx \omega_L$ and large $g_c/\gamma \gg 1$. \textbf{(c, d)} Cavity occupations corresponding to the same parameters as in \textbf{(a, b)}. \textbf{(e)} Minimal $g^{(2)}(0)$ considered over wide detuning $\Delta$ range, plotted as a function of $g_c$, showing both unconventional and conventional blockade regions for various electron numbers $N_s$. \textbf{(f)} Time-delayed second order coherence $g^{(2)}(\tau)$ where solid curves depict results for $g_c = 1.2\gamma$ at optimal detuning and increasing $N_s$. Dashed curve corresponds to the case of conventional blockade ($g_c=120\gamma$ and $N_s=100$).}
\label{fig:g2}
\end{figure}
The crucial point of derivation corresponds to the off-diagonal elements, coming from strong light-matter coupling Hamiltonian $\hat{\mathcal{H}}_{\mathrm{coupl}}$, which read [SM]
\begin{align}
\label{eq:off-diag}
\notag
    &\mathcal{M}_{N_C, N_T -1}^{N_C - 1, N_T} = \frac{\Omega}{2} \sqrt{N_C N_T \left( 1 - \frac{N_T}{N_s + 1}\right)} \sqrt{\frac{N_s}{N_s + 1}} \\ & \times \left[ 1 - (-1)^{N_T} \frac{(N_s - N_T)! N_T!}{N_s!}\right].
\end{align}
Importantly, expression \eqref{eq:off-diag}: 1)  holds for arbitrary $N_T \leq N_s$; 2) is valid for highly nonlinear case of $N_s = 1$ (corresponding to a qubit) \cite{Agranovich}; 3) provides physical result for singly occupied mode $N_T = 1$, unlike for Holstein-Primakoff approach \cite{Holstein,Emary2003,Brandes2003} widely used for large $N_T$ but failing in this limit. This is crucial for calculating the quantum statistical properties of trion-polaritons.

\textit{Trion-polariton spectrum.---}Once blocks $\hat{H}_N$ are known, the polariton energies can be found separately for each value of $N$. For concreteness, we assume the monomode approximation corresponding an effectively 0D open microcavity \cite{Dufferwiel2017}, such that only zero momenta photons are considered, $\hat{c}:= \hat{c}_{\mathbf{k}\rightarrow 0}$, $\hat{B}:=\hat{B}_{\mathbf{k}\rightarrow 0}$, $\hat{H}_{coupl}:=(\Omega/2)\left(\hat{B}^\dagger\hat{c}+\hat{B}\hat{c}^\dagger\right)$. We consider the system driven by a strong coherent pulse, such that initial particle distribution corresponds to a coherent state for the photonic field given by the Poisson distribution with amplitude $\alpha$, while the trion mode remains unoccupied. We calculate the corresponding transmission spectrum $S(\omega)$ (SM, section B) for increasing total number of photons $ N_C=|\alpha|^2$. The cavity output, modified by the strong coupling to trions, is then monitored in the transmission geometry. The results are shown in Fig.~\ref{fig:spectrum}(a), where we considered zero cavity-trion detuning $\delta = \omega_{\mathrm{cav}} - \omega_T$, fixed dissipation rates of $\gamma_c = \gamma_T = 0.05 \Omega$, and $N_s = 100$ available electronic states. For small $N_C$ the two trion-polariton peaks are clearly visible, representing the expected Rabi doublet. As $N_C$ increases the number of transitions grows and the distance between the peaks decreases [see Eq.~\eqref{eq:off-diag}]. Finally, at large occupations $N_C > 100$ the two peaks merge and the broad band of transitions is visible in the spectrum. To track the collapse of the strong coupling, we plot peak positions as a function of pump ($N_C$) for different dissipation rate [Fig.~\ref{fig:spectrum}(b)]. At large decay rates ($\gamma/\Omega = 0.1$) the collapse is shifted to smaller $N_C$ values, while for narrow lines it saturates at $N_C = N_s$.

\textit{Phase space filling induced antibunching.---}To calculate the quantum statistics for the cavity field we consider the finite Hilbert space with matrix elements modified due to the phase space filling as shown in Eq.~\eqref{eq:off-diag}. We consider the case of weak coherent cw pump with frequency $\omega_p$ and strength $P$ detuned by $\Delta=\omega_{\mathrm{cav}}-\omega_p$ from the energy of the cavity mode. The dynamics for the system is studied using the master equation approach, where Lindblad dissipation terms with collapse operators $\sqrt{\gamma_c} \hat{c}$ and $\sqrt{\gamma_T} \hat{B}$ are introduced. As we focus on the regime of few quanta $N_C$ and $N_T$ are truncated in the way that higher states are negligibly populated (we consider the range of pumps for which $N_{C}, N_{T}<10$).

To characterize the statistics of the cavity output we calculate the second-order coherence function at finite time delay $\tau$, $g^{(2)}(\tau) = \langle \hat{c}^\dagger (0) \hat{c}^\dagger (\tau) \hat{c} (\tau) \hat{c}(0) \rangle/\langle \hat{c}^\dagger \hat{c} \rangle^2 $, as well as steady state intracavity occupation $n_{\mathrm{cav}} = \langle \hat{c}^\dagger \hat{c} \rangle$. The results are shown in Fig.~\ref{fig:g2}, where for brevity we concentrate on the case of equal decay rates $\gamma_c = \gamma_T =: \gamma$ and zero trion-photon detuning $\omega_{\mathrm{cav}} = \omega_T$ (see SM for full characterization). 
Studying the dependence of second-order coherence at zero delay $g^{(2)}(0)$ on the pump detuning $\Delta$ we reveal two qualitatively different regimes of antibunching coming from the optical saturation of the trion-photon coupling. At small single trion coupling $g_c=\Omega/\sqrt{N_s}$ being comparable with cavity and trion linewidth we find the region of pronounced antibunching which can be attributed to the unconventional photon blockade \cite{Liew2010,Bamba2011}, where due to destructive interference the two-photon occupation vanishes [Fig.~\ref{fig:g2}(a)]. Namely, there are two excitation paths to populate two-photon state---one from direct coherent excitation, and second via the alternative root through the trion mode [Fig.~\ref{fig:sketch}(b)]. For certain optimal conditions the two interfere destructively, leading to largely reduced two-photon probability. The process requires pump to be nearly resonant with the cavity mode and $g_c/\gamma \sim 1$, and notably the optimal pump position does not depend on $g_c$. Already at $g_c/\gamma = 2$ the $\Delta \approx 0$ antibunching window disappears, and instead the single photon emission at lower polariton frequency $\omega_p \approx \omega_L = (\omega_{\mathrm{cav}} + \omega_T)/2 - \sqrt{\Omega^2 + \delta^2}/2$ emerges. In Fig.~\ref{fig:g2}(b) this corresponds to minimum at $\Delta/\gamma \approx \Omega/2$ which shifts with $g_c$. The Fano-lineshape profile of $g^{(2)}(0)$ [Fig.~\ref{fig:g2}(b)] and requirement of the strong single trion-photon coupling $g_c/\gamma \gg 1$ allows to attribute it to conventional blockade, comparable to antibunching in Kerr-type nonlinear systems \cite{Verger2006}. The two quantum regimes can be also characterized by the probability of the single photon emission being proportional to cavity occupation $n_{\mathrm{cav}}$ plotted in Fig.~\ref{fig:g2}(c,d). As the unconventional antibunching window lies in the middle of the polaritonic spectra, the associated occupations lie in $10^{-3}$..$10^{-4}$ range [Fig.~\ref{fig:g2}(c)], for relevant $g_c/\gamma \sim 0.5-1.2$ values ($P/\gamma = 0.5$ is considered). For $\omega_L$-resonant pump the occupation peaks at $\sim 10^{-1}$ values simultaneously with minimal $g^{(2)}(0)$ [Fig.~\ref{fig:g2}(d)], at the expense of weaker antibunching and large $g_c/\gamma$ requirement. The performance in both regimes can be further characterized by plotting minimized $g^{(2)}(0)$ for both detuning windows as a function of light-matter coupling $g_c$ and number of electrons $N_s$ [Fig.~\ref{fig:g2}(e)]. While in the unconventional regime low $g^{(2)}(0)$ holds well for $g_c/\gamma < 2$ and does not depend on $N_s$, the conventional case for $g_c/\gamma > 2$ shows $N_s$ dependence where low electron concentration is favored.
Finally, the important dependence of the single photon emission is finite delay response, which defines how well the emitted single photon can be resolved \cite{Bayer2010}. As expected for the interference effect, the unconventional trion-polariton blockade plotted for optimal $\Delta$ and $g_c/\gamma = 1.2$ shows oscillations in $g^{(2)}(\tau)$ with period inversely proportional to $\sqrt{N_s}$, where the antibunching region shrinks as $N_s$ grows [solid curves in Fig.~\ref{fig:g2}(f)], while remaining a significant portion of $\gamma^{-1}$ even for large occupation $N_s = 100$. We compare it to the conventional blockade at two-orders larger coupling $g_c = 120\gamma$ ($N_s = 100$), which does not show oscillations, yet increases with $\tau$.

\textit{Discussion.---}To get the quantitative estimates for trion-based antibunching in TMD materials, the characteristic strength of light-matter coupling between a cavity photon and a trion can be estimated as $g_c = g_0\chi_T$, where $g_0$ is the bare e-h coupling constant, and $\chi_T$ is a trion confinement coefficient coming from integrating the relative motion wavefunction. We adopt the approach from Refs. \cite{Combescot2003b,Shiau2012,Combescot2009,Shiau2017} and consider the standard Chandrasekhar-type wavefunction for the trion with two variational parameters \cite{Ramon2003}, which was shown to work well for nearly equal electron/hole effective masses \cite{Courtade2017}. In this case, the confinement coefficient becomes
\begin{equation}
\chi_T = [8 (\lambda_1^2 + \lambda_2^2)^2 (\lambda_1 + \lambda_2)^4 / \{ \lambda_1^2 \lambda_2^2 (\lambda_1 + \lambda_2)^4 + 16 \lambda_1^4 \lambda_2^4 \} ]^{1/2},
\end{equation}
with $\lambda_1$ and $\lambda_2$ being variational parameters corresponding to an effective radii of electrons in a trion (can be understood as exciton-like shell and outer electron shell properties). Considering $\lambda_2 > \lambda_1$, the limit of $\lambda_2/\lambda_1 \gg 1$ is favored for achieving large $\chi_T$. The bare electron-hole pair coupling can be calculated as \cite{HaugKoch}
\begin{equation}
\label{eq:g0}
 g_0 = e p_{cv} \sqrt{\xi^2 \hbar^2 / 2\epsilon \epsilon_0 m_0^2 \omega_0 L_{\mathrm{cav}} A}=\sqrt{\xi^2 \hbar^2 e^2/\epsilon \epsilon_0 \mu L_{\mathrm{cav}} A},
  \end{equation}
 where $e$ is an electron charge, $p_{cv}$ is an interband transition matrix element, $\xi$ accounts for TMD placement in the cavity ($\xi = 1$ corresponds to an antinode), $\epsilon$ is dielectric medium permittivity, $\epsilon_0$ is vacuum permittivity, $m_0$ is a free electron mass, $\omega_0$ is a transition energy, $L_{\mathrm{cav}}$ is a cavity length, $\mu = (1/m_e + 1/m_h)^{-1}$ is the reduced electron-hole pair mass (measured in units of $m_0$), and $A$ is the area of the sample.

As a particular example we consider a MoSe$_2$ flake inside an open cavity. The parameters of a standard setup are chosen as effective cavity length $L_{\mathrm{cav}} = 1~\mu$m, cavity area of $A=1~\mu$m$^2$, electron density of $10^{10}$~cm$^{-2}$, and optical linewidth of $\gamma_{c} = 50~\mu$eV. The effective masses of electron and hole in MoSe$_2$ $m_e = 0.8$ and $m_h = 0.84$ \cite{Larentis2018}, and Eq.~\eqref{eq:g0} gives $g_0 = 0.058$~meV ($\xi=1$), as expected for a direct bandgap semiconductor \cite{NoteCoupling}. Performing the variational procedure for the relevant case of MoSe$_2$ on hexagonal boron nitride (hBN) the trion radii are $\lambda_1 = 0.87$~nm and $\lambda_2 = 2.54$~nm, providing the enhancement coefficient of $\chi_T = 7.35$. The nonradiative decay rate for trions was measured in hBN encapsulated samples $\gamma_T = 0.26$~meV due to inhomogeneous exciton broadening \cite{Martin2018}. Considering the TMD monolayer placed outside of the antinode with $\xi = 0.6$, yielding $g_c = 0.256$~meV, the setup can provide unconventional antibunching of $g^{(2)}_{\mathrm{MoSe}_2}(0) = 0.064$ with $n_{\mathrm{cav}} = 0.00013$.  
Improving the system, trion non-radiative decay can be reduced to $\gamma_T\sim 10~\mu$eV at $T=1$~K temperature given by phonon interactions only \cite{Martin2018}, and we note that the cavity linewidth $\gamma_{c} \sim 10~\mu$eV was already realized in the state-of-the-art setups \cite{Besga2015,Dufferwiel2015c}. Thus, for optimally coupled layer ($\xi=1$) the conventional antibunching value of $g^{(2)}_{\mathrm{MoSe}_2}(0) = 0.091$ can be obtained. One can compare this to the potential state-of-the-art GaAs sample with a $2.3\mu$m diameter micropillar cavity of the same effective area, where exciton-exciton interaction-based Kerr blockade \cite{Verger2006} can give $g^{(2)}_{\mathrm{GaAs}}(0) = 0.92$ for the same values of broadening, the limit nearly approached experimentally, where second order coherence at 0.95 level was reported \cite{Munoz-Matunano2019,Delteil2019}. 


\textit{Conclusions.---}We developed a theory of quantum nonlinear optical response of trion-polaritons fully accounting for their composite nature and related phase-space filling effects up to infinite order. Analyzing the transmission spectrum of the system, we observed and described quantitatively the collapse of the strong light-matter coupling with increase of the optical pump. We studied the effects of quantum correlations in the system, and revealed the rich phenomenology where both unconventional and conventional blockade can be studied in regimes of weak and strong single trion-photon coupling, correspondingly. We found that strong antibunching of the photonic emission is possible with TMD monolayers put in an open microcavity, being accessible in modern and near-term setups. The results offer a new vista for development of quantum polaritonics in the planar samples without electronic confinement.

\begin{acknowledgments}
\textit{Acknowledgments.---}We would like to thank Alexey Kavokin and Ata\c{c} Imamoglu for the useful discussions. The work was supported by the Government of the Russian Federation through the Megagrant 14.Y26.31.0015, ITMO Fellowship and Professorship Program, and Russian Science Foundation Project No. 18-72-10110.
\end{acknowledgments}


\end{document}